\documentclass[fleqn,usenatbib]{mnras}

\usepackage[T1]{fontenc}

\DeclareRobustCommand{\VAN}[3]{#2}
\let\VANthebibliography\thebibliography
\def\thebibliography{\DeclareRobustCommand{\VAN}[3]{##3}\VANthebibliography}
\def\tc{t_{\rm cr}}
\def\tr{t_{\rm r}}
\def\torb{t_{\rm orb}}
\def\tage{t_{\rm age}}
\def\sigmaT{\sigma_{\rm T}}
\def\erf{{\rm erf}}

\usepackage{graphicx}	% Including figure files
\usepackage{amsmath}	% Advanced maths commands
\usepackage{amssymb}	% Extra maths symbols
\usepackage[dvipsnames]{xcolor}

\newcommand{\gaia}{{\it Gaia}}
\newcommand{\bprp}{\ensuremath{\mathrm{BP}-\mathrm{RP}}}
\newcommand{\kms}{\ensuremath{\mathrm{km}\,\mathrm{s}^{-1}}}
\newcommand{\wyn}[1]{{\textcolor{black}{#1}}}
\newcommand{\so}[1]{{\textcolor{black}{#1}}}

\newcommand{\rh}{r_{\rm h}}
\newcommand{\Xcrit}{X_{\rm crit}}
\title[Mass Segregation in the Hyades]{Mass Segregation in the Hyades Cluster}

\author[Evans \& Oh]{
	N. Wyn Evans$^{1}$\thanks{Email: nwe,soh@ast.cam.ac.uk} and Semyeong Oh$^{1}$
	\\
	$^{1}$Institute of Astronomy, University of Cambridge, Madingley Rd, Cambridge, CB3 0HA, UK\\
}

\date{Accepted XXX. Received YYY; in original form ZZZ}

\pubyear{2021}

\begin{document}
\label{firstpage}
\pagerange{\pageref{firstpage}--\pageref{lastpage}}
\maketitle

\begin{abstract}
%We study the extent and cause of mass segregation in the Hyades. 
Using the \gaia\ colour-magnitude diagram, we assign masses to a catalogue of 979 confirmed members of the Hyades cluster and tails. By fitting the cumulative mass profile, stars within the tidal radius have a Plummer-like profile with half-mass radius $r_{\rm h}$ of 5.75 pc. The tails are extended with $r_{\rm h} = 69.35$ pc and fall off more slowly than Plummer with density proportional to distance$^{-1.36}$. The cluster stars are separated into two groups at BP-RP $=2$ or $0.56 M_\odot$ to give a high mass (${\bar M} = 0.95 M_\odot$) and a low mass (${\bar M} = 0.32 M_\odot$) population. We show that: (i) the high mass population has a half-mass radius $r_{\rm h}$ of 4.88 pc, whilst the low mass population has $r_{\rm h} = 8.10$ pc; (ii) despite the differences in spatial extent, the kinematics and binarity properties of the high and low mass populations are similar. They have isotropic velocity ellipsoids with mean 1d velocity dispersions $\sigma$ of 0.427 and 0.415 km\,s$^{-1}$ respectively. The dynamical state of the Hyades is far from energy equipartition ($\sigma \propto {\bar M}^{-1/2}$). We identify a new mass segregation instability for clusters with escape speed $V$. Populations with $V/\sigma \lesssim 2\sqrt{2}$ can never attain thermal equilibrium and equipartition. This regime encompasses many Galactic open and globular clusters. For the Hyades, there must be an outward energy flux of at least $9.5 \times 10^{-4} M_\odot\,{\rm km^2\, s^{-2} Myr^{-1}}$ to maintain its current configuration. The present mass loss of $0.26 M_\odot {\rm Myr}^{-1}$ due to tidal stripping by itself implies a substantial energy flow beyond the required magnitude.
\end{abstract}


%%%%%%%%%%%%%%%%%%%%%%%%%%%%%%%%%%%%%%%%%%%%%%%%%%

%%%%%%%%%%%%%%%%% BODY OF PAPER %%%%%%%%%%%%%%%%%%

\section{Introduction}

The metamorphosis of open clusters is driven by interplay between external and internal processes. Even for isolated clusters, there are evolutionary effects such as two-body relaxation, mass segregation, collisions, stellar evolution and the hardening of binaries. External dynamical disturbances — tidal stripping by the Galactic gravitational field, disc shocking, encounters with Giant Molecular Clouds — can influence equally strongly the dynamical evolution of open clusters~\citep[e.g.][]{Da15,Me21}. Although seemingly simple objects of at most a few thousand stars, the dynamical state of open clusters can be tangled and convoluted to unravel.

The internal cluster dynamics is controlled by (at least) three timescales. They are the crossing time $\tc$, which is the time taken by a star to move across the system; the two-body relaxation time $\tr$, which is the time needed by stellar encounters to redistribute the energies of stars; and the orbital period $\torb$, which is the time during which external tidal effects become significant.

The Hyades is the nearest open cluster to us ($d \approx 46$ pc) and is an obvious target for studies of the internal kinematics~\citep{Pe98, deB01}. For Hyades, the crossing time is $\tc \approx 50$ Myr, which is comparable to the relaxation time $\tr \approx 60$ Myr ~\cite[][hereafter OE]{Oh20}. The orbital period is $\torb \approx 200$ Myr \citep{Er11}, so the Hyades has $\tc \approx \tr < \torb$. By comparison, the age of the Hyades is $\tage \approx 680$ Myr \citep[e.g.,][]{Br15,gossage2018,lod18}. Therefore, the Hyades is expected to show both tidal tails ($\torb < \tage)$ and mass segregation ($\tr < \tage$).

Early numerical simulations predicted the existence of Hyades tidal tails~\citep{Ch05,Er11}. The actual discovery of the tails~\citep{Me19,Ro19} followed quickly after the second \gaia\ data release~\citep[DR2,][]{gaia-dr2-release}, and was in reasonable accord with the predictions. Using DR2, the leading tail extends up to 170 pc, and the trailing tail up to 70 pc, with the origin of the asymmetry unclear. With the advantage of \gaia\ early data release 3 (eDR3), \citet{Je21} found that the tails extend to nearly 1 kpc in total! The enormity of the Hyades tidal tails already tells a story of very substantial mass loss. OE showed that the cluster has a 'super-virial' velocity dispersion, consistent with disruption. From simple models of the mass loss, they estimated that the Hyades is in its death throes with only a further $\le$ 30 Myr left to live. Most of the original mass of the Hyades has already been lost.

Evidence for mass segregation in the Hyades dates back to at least \citet{Re92} and \citet{Eg93}, who both noted the abundance of bright stars and the deficiency of faint ones in the central regions, concluding that the luminosity function is different from the field population. Using {\it Hipparcos} data, \citet{Pe98} observed that the less massive stars are more spatially extended than the more massive ones. This was subsequently confirmed with PanSTARRS data by \citet{Go13} and with \gaia\ DR2 by~\citet[][see also Fig. 8 of OE]{Lo19}. The phenomenon is immediately obvious in \gaia\ DR2, for example, by plotting the colour-magnitude diagram as a function of radius. However, mass segregation in the Hyades has not so far been studied quantitatively, despite the fact that the richness of the \gaia\ datasets allow for the spatial, kinematic and binarity properties of the mass-segregated populations to be scrutinised in some detail. Our paper rectifies this omission.

Mass segregation can have a primordial origin in youthful star clusters~\citep[e.g.,][]{Bo98,Mc07}. The age and end-of-life status of the Hyades however argues strongly that the prominent mass segregation is a consequence of two-body relaxation and tidal stripping. The time scale for dynamical friction to substantially decrease the energy of a massive star of mass $m_1$ is less than the relaxation time scale for lighter stars of mass $m_2$ by a factor $m_2/m_1$ ~\citep[see eq 14.65 of][]{Sa85}. As massive stars lose energy to lighter ones, they sink to the centre of the cluster. On the other hand, the lighter stars on average increase their total energy and move into the outermost parts of the cluster, or even escape to form the tails ~\citep[e.g., Section 7.5.5 in][]{BT}. By analogy with the kinetic theory of gases, we might expect energy equipartition to be set up with $m_1\sigma_1^2 = m_2\sigma_2^2$, where $\sigma_1$ and $\sigma_2$ are the respective velocity dispersions.

However, the congregation of the high mass stars in the centre causes a deepening of the cluster potential. So these stars experience a higher binding energy and, via the virial theorem, they begin to move faster. This is one of the paradoxical consequences of the negative heat capacity of self-gravitating systems~\citep[e.g.,][]{LBW}. If this process runs to completion, the endpoint is core contraction with an infinite density~\citep{LBE80}. In reality, a number of processes intervene. Binary stars can form and harden by interaction with other stars, thus absorbing the negative energy that continues to flow into the centre~\citep[e.g.,][]{Kr20}. Open clusters are moving on (usually mildly) eccentric orbits, and the time-varying Galactic tidal field strips mass and consequently energy as well. Tidal shocks produced by passing Giant Molecular Clouds and disk crossings also can act as a disruptive presence. Both these effects disrupt binaries within the cluster, as well as the cluster itself.

Another way of understanding the absence of energy equipartition was provided by \citet{Sp69}, who studied analytically the equilibrium between high and low mass populations with Maxwellian distributions. If the total mass in the heavier population is too great, then it is not possible for the systems to be in both dynamical and thermal equilibrium together. Equipartition with $\sigma \propto m^{-1/2}$ then cannot be attained, because the subsystem of more massive stars decouples and becomes nearly self-gravitating. This insight was confirmed by later calculations~\citep{Vi78} and numerical experiments~\citep{Tr13, Pa16}. So, even in the absence of external effects -- always important for such feeble entities as open clusters -- equipartition need never be attained.

There are a limited number of simulations of the collisional evolution of open clusters with multi-mass components and stellar evolution in (usually) static Galactic tidal field in the literature. \citet{Sp16} found that such open clusters become strongly mass segregated, even though the velocity dispersion remains almost flat as a function of mass. Hard binaries start to form and prevent core collapse at $t \approx 0.5 \tr$ in these simulations. At this stage, the velocity dispersion of the most massive stars ($\gtrsim 10\, M_\odot$) becomes higher than the velocity dispersion of the lighter stars. In other words, the open clusters remain far from equipartition with scant dependence of kinematics on stellar mass. Although interesting, the open cluster simulations in \citet{Sp16} are not especially good matches to the Hyades today. In particular, the masses of the stars at the endpoint of the simulations extend above $20 M_\odot$, whilst the heaviest stars in our Hyades catalogue are $\sim 2.5 M_\odot$. The simulations are only run to $\approx 6$ relaxation times, less than the present-day state of the Hyades. Also, the static tidal field does not do as much damage as a time-varying tidal field when the cluster moves on an eccentric orbit.

\citet{Er11} provide simulations tailored to the Hyades and incorporating collisional and stellar evolution. They find reasonable agreement between their models and the pre-\gaia\ data on mass segregation, as judged by a minimum spanning tree statistic. They suggest that the Hyades is a moderately mass-segregated model star cluster. Very recently, \citet{Je21} presented the N-body evolution of a Hyades cluster on a realistic orbit in the Milky Way galaxy. They do not directly address mass segregation, but do suggests that the asymmetry of the tails and the 'super-virial' velocity dispersion \citep[][see also OE]{Ro11} can both be explained by a recent interaction with a massive Galactic perturber. This substantiates the suggestion of OE that Hyades is far from equilibrium and probably in its death throes.

Here, we quantify the mass segregation in the Hyades using the improvements brought by the \gaia\ photometry and astrometry. This gives us the opportunity to examine the spatial distribution, 3-d kinematics and binarity as a function of mass of the population for the first time. In Section 2, we associate masses with confirmed Hyades members and so divide the cluster into high and low mass populations. 
By fitting the cumulative mass profiles, we show that the entire cluster has a half-mass radius $\rh \approx  5.75$ pc, somewhat larger than earlier studies~\citep[e.g.,][]{Pe98}. The half-mass radii of the high mass and low mass populations are 4.88 pc and 8.10 pc. Despite strong mass segregation, the two populations are barely distinguishable kinematically. Both have almost isotropic velocity dispersions $\sigma \approx 0.4$ kms$^{-1}$. The combination of radial velocity variability and \gaia's renormalised unit weight error (RUWE) suggest that the binary properties of the two populations are similar. 

Section 3 then develops models of the thermal equilibrium of two populations in a cluster with finite escape speed $V$. We show the existence of a new mass segregation instability when $\sigma/V \approx 3$, which prevents energy equipartition. The Hyades cluster -- together with many of the Galactic open and globular clusters -- fall into this regime in which energy equipartition is unattainable. The difference in the half-mass radii of the two Hyades populations is maintained by heat flow through the cluster, which manifests itself as the mass loss.

\section{Data Analysis}
\label{sec:Analysis}

\begin{figure}
    \centering
    \includegraphics[width=\linewidth]{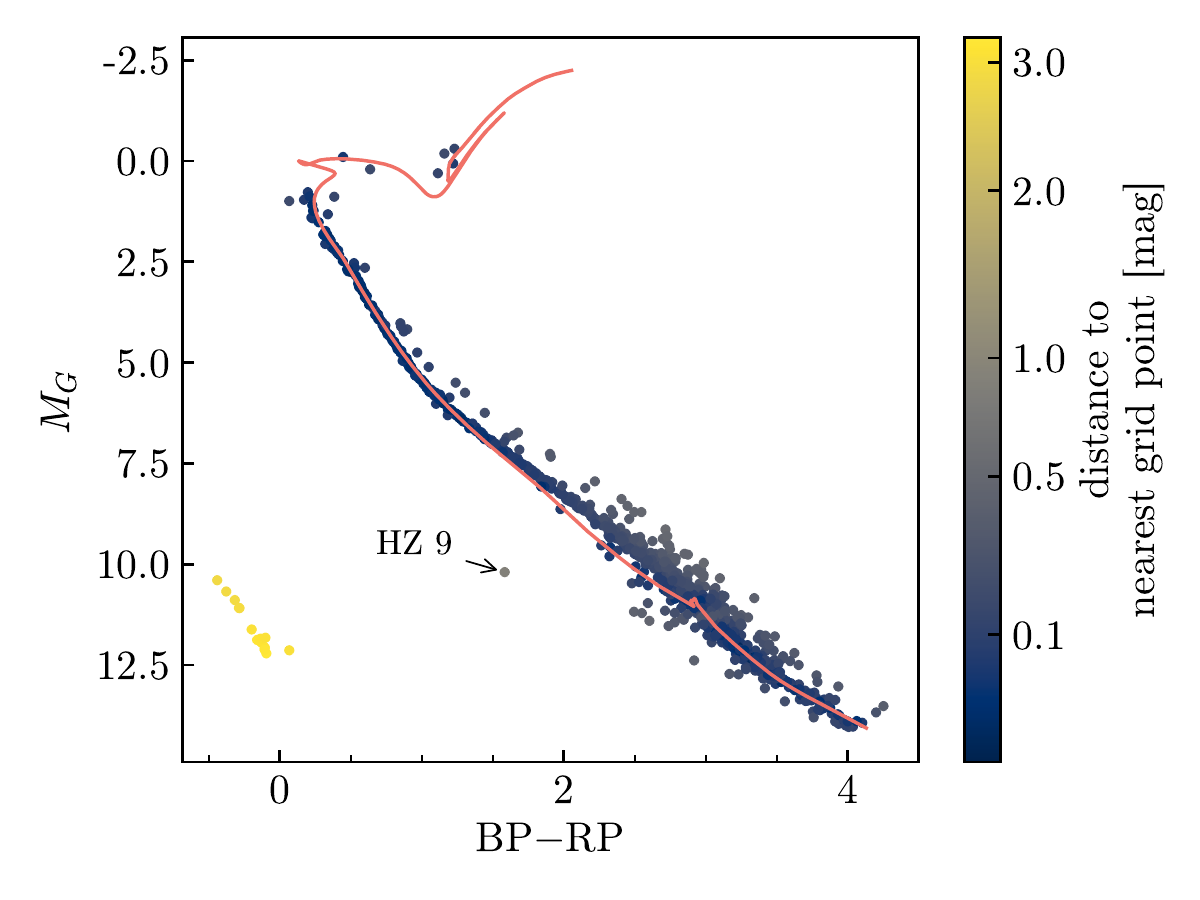}
    \caption{Assigning stellar mass to each source in \gaia's colour-magnitude space. We overplot the MIST isochrone used to assign mass to each source on top of the data. The data are coloured by distance to the nearest grid point in magnitude. Because the white dwarf sequence is not covered by the isochrone model, they stand out in colour and the mass assigned can be wrong. The region between the main sequence and the white dwarf sequence can be occupied by unresolved WD-MS binaries. One data point that clearly falls in this region corresponds to HZ 9, a known post-common envelope spectroscopic binary \citep{La81}.
    }
    \label{fig:cmd}
\end{figure}

\begin{figure*}
    \centering
    \includegraphics[width=\linewidth]{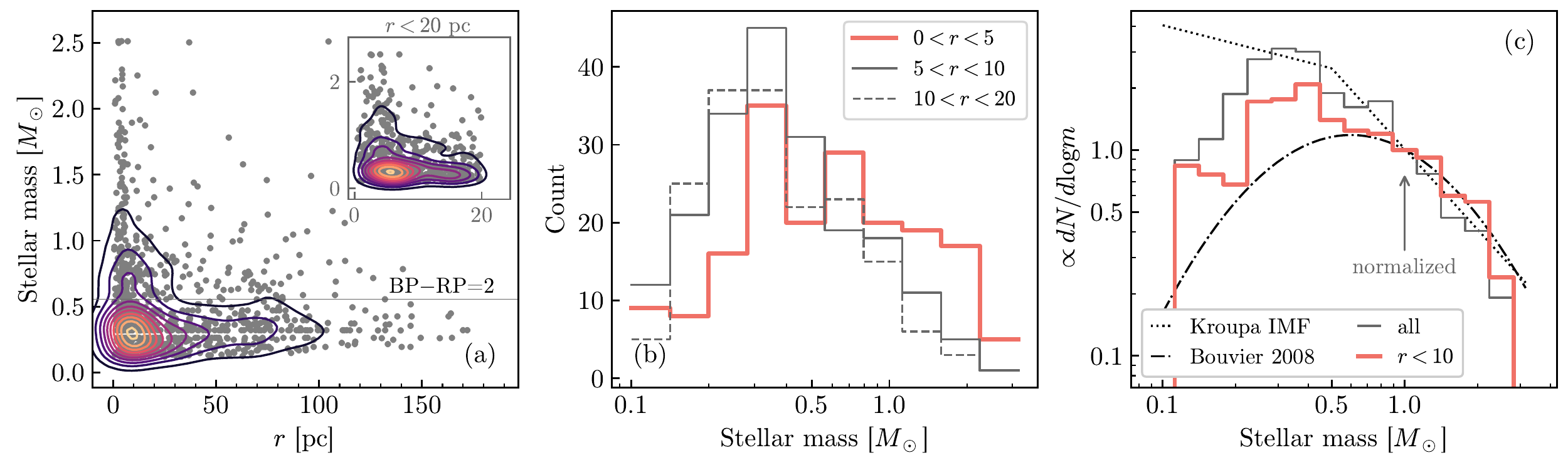}
    \caption{Mass segregation in the Hyades cluster.
        (a) Sample distribution on the stellar mass vs. radius plane.
        The contours are generated from 2D kernel density estimation using a Gaussian kernel. The inset axes shows the same plot but only using stars within $20$~pc (twice the tidal radius). More massive stars are more centrally-concentrated, and the tidal tails are preferentially comprised of lower mass stars. The horizontal line indicates the stellar mass corresponding to our colour cut of $\bprp=2$ dividing ``high-mass'' and ``low-mass'' sample.
        (b) Histogram of stellar mass in different radius bins.
        The central $5$~pc shows a distinctively different distribution skewed towards higher mass while
        the difference between $5<r<10$ and $10<r<20$ is more subtle.
        (c) Mass function for the entire sample (black) and the stars within $10$~pc (red).
        We compare this to the Kroupa initial mass function and the best-fit present-day mass function of
        the Hyades cluster from \citet{bouvier2008}.
        All mass functions are normalized at $1~M_\odot$.
    }
    \label{fig:massseg}
\end{figure*}

\subsection{Mass Assignment}
\label{sub:mass-assignment}

We use the \gaia\ DR2 \citep{gaia-dr2-release} astrometric and photometric data for the Hyades cluster and tail stars presented in OE. There, we developed a two component mixture model to distinguish cluster and tail members from background contaminants while simultaneously modelling the internal kinematics of the cluster. We provided a sample consisting of 1103 Hyades candidates in Table 2 of OE, together with a kinematic membership probability. In what follows, we exclude a small number of sources that have membership probability below $0.5$.

We assign mass to each source by nearest-neighbour interpolation on the \gaia\ colour-magnitude space ($\bprp$ vs. $M_G$) using the MIST isochrone \citep{choi2016-mist} of the age and metallicity of the Hyades cluster:
$[\mathrm{Fe}/\mathrm{H}] = 0.24$ and $\log \mathrm{age} = 8.83$ \citep{gossage2018}. We only consider sources with absolute $G$-band magnitude brighter than the MIST isochrone model grid, $M_G < 14.06$, which corresponds to stellar mass larger than $\approx 0.12 M_\odot$. This additional restriction reduces the sample to 979 sources.

Fig.~\ref{fig:cmd} shows the data and the isochrone in \gaia\ colour-magnitude space. The coPosterlour of the data points indicate distance to the nearest grid point, which reveals the limitation of our mass estimation: sources that deviate from the single-star main sequence (MS) such as white dwarfs (WDs) and unresolved binaries of either two MS stars (above the MS) or MS-WD (in between the MS and WD sequence, such as the known post-common envelope binary HZ~9 ~\citep{La81}) end up with inaccurate masses. The 50\% and 90\% percentiles of the distances to the nearest grid point is $0.092$ and $0.26$~mag. We examine the effect of incorrectly assigned mass for binaries, which may be off by at most a factor of $\approx 2$, later in Section~\ref{sub:effects-of-binaries}. It is worth noting that, even for the sources with a small distance to the nearest grid point, the distribution of distances (in magnitudes) in units of the propagated measurement uncertainties are systematically offset from 0. This underlines the fact, though we need to rely on isochrone models to assign stellar mass given observable properties, they are not perfect even for single MS stars.

The tidal radius provides a crude division of the cluster proper from the tails. It is taken as $r_{\rm t} \approx 10$~pc from \citet{Ro11}, where it is found by equating the mean density within $r_{\rm t}$ to the mean density within the cluster's orbit, assumed circular (see their Figure 6). This is similar to other literature values~\citep{Pe98,Er11}. The tidal radius is imprecisely defined for satellites on eccentric orbits~\citep{Re06}, so further refinement of its value is not really warranted.

The left panel of Fig.~\ref{fig:massseg} shows the stellar mass as a function of cluster-centric distance out to 200~pc, well beyond the tidal radius. The effects of mass segregation are clearly visible in the contours. The inset figure shows that the high mass stars are preferentially located at small cluster-centric distance within $r<20$~pc.
%Stellar mass and cluster-centric distance anti-correlates within the cluster (see the inset figure). 
This trend continues beyond the tidal radius as stars of lower mass are also more likely to leave the cluster, populating the tidal tails.

We examine how the mass function changes as a function of radius in the middle panel of Fig.~\ref{fig:massseg}. The mass function of the central part ($0<r<5$) of the cluster is highly skewed towards higher mass, while the change between $5<r<10$ and $10<r<20$ is more subtle. Using the two-sample Kolmogorov-Smirnov test, we can reject the null hypothesis that the sample for $0<r<5$ comes from the same distribution as those from the other two radius bins ($p$-value less than $10^{-4}$). Between $5<r<10$ and $10<r<20$, we cannot reject the null hypothesis.

Finally, we compare the mass function of the entire sample and stars within the tidal radius ($10$~pc) to the Kroupa initial mass function \citep[IMF;][]{kroupa2001} in the right panel of Fig.~\ref{fig:massseg}. In order to make relative comparisons, we normalize all mass functions on this plot at $1~M_\odot$. The mass functions for both samples are distinctively different from the Kroupa IMF. Specifically, both samples contain less low-mass stars than naively expected from the nominal Kroupa IMF, and the deviation starts at higher mass for the cluster sample.
We note that while we do not make any attempt to correct for completeness of this sample selection, the lowest mass considered here corresponds to $G\approx 17.5$~mag at the distance of the cluster, which is well above where the magnitude at which \gaia's survey selection function becomes significantly incomplete \citep[$G\approx19$;][]{boubert2020}. As for the incompleteness induced by the astrophysical sample selection such as data quality cuts and kinematic membership selection \citep{rix2021}, we think that this should still be minimal for the cluster sample ($r<10$~pc) given that the cluster is very nearby and distinct in its kinematics. However, the selection of tidal tail stars is less trivial and may suffer from more significant and complex incompleteness.  As we focus on the kinematics of cluster stars in this study, we have not explored further on this issue.

The lack of low-mass stars in Hyades has already been noted by e.g., \citet{bouvier2008},
who attributed the deficiency to preferential loss of low-mass stars due to dynamical evolution. We compare our mass function to their best-fit in Fig.~\ref{fig:massseg}c, and find that the deficiency is not as large in our sample. The difference may be due to the extended membership list made possible by the \gaia\ data.

Current methods to quantify mass segregation fall into two
categories~\citep[e.g.,][]{Al09}. We can either fit a density profile and
characteristic radii to various mass ranges. Or we can bin in radii and trace
the variation of the mass function with radius. As the \gaia\ selection function
depends on apparent magnitude but does not depend on cluster-centric radius, we
adopt the former here. Retaining the sources within the tidal radius, we divide
the sample into ``high-mass'' and ``low-mass'' bins with a colour cut $\bprp=2$,
which corresponds to $0.56 M_\odot$. The number of sources in the high-mass and
low-mass bins are 144 and 231 respectively. The median component mass for
high-mass and low-mass bins are $0.95 M_\odot$ and $0.32 M_\odot$.

\subsection{Cumulative Mass Profiles}
\label{sub:cumulative-mass-profiles}

\begin{figure*}
    \centering
    \includegraphics[width=0.95\linewidth]{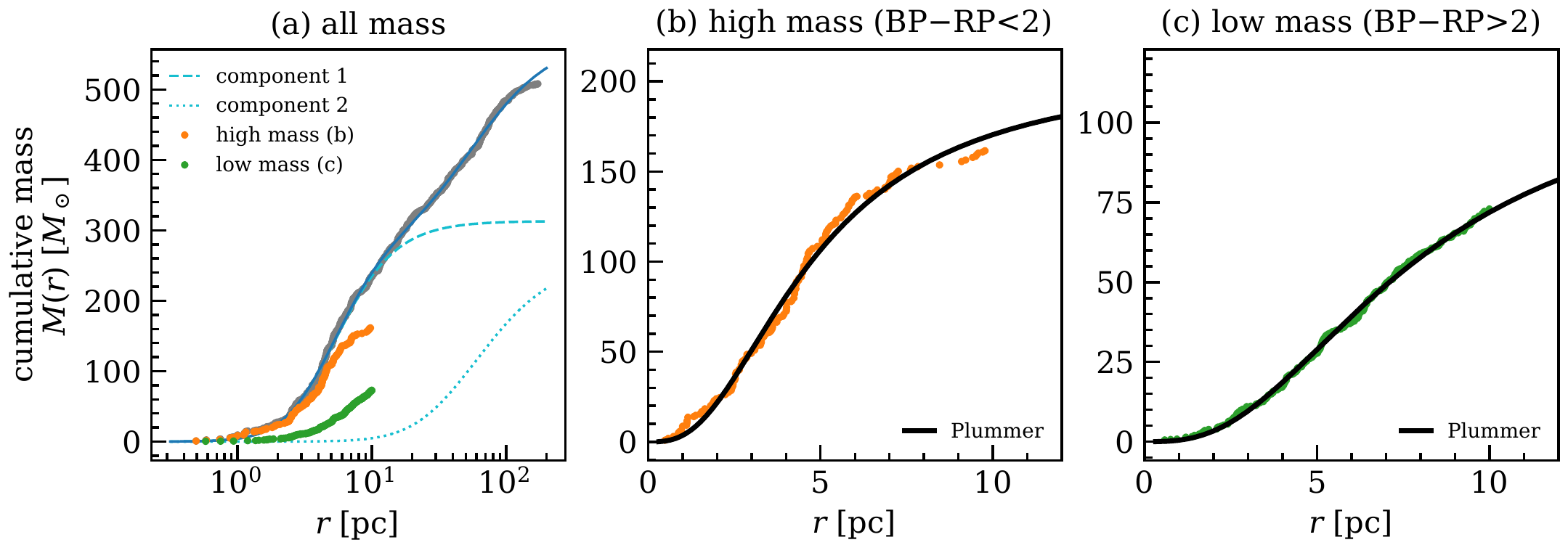}
    \caption{Fitting cumulative mass profile of all stars (a),
        high-mass (b) and low-mass (c) stars within the cluster ($r_c<10$~pc). In each panel, the data are plotted as points and the best-fit models (and their components) are plotted as lines (see Section~\ref{sub:cumulative-mass-profiles} for details). Best-fit parameters are summarized in Table~\ref{tab:fit-result}.
    }
    \label{fig:cumulative-mass-fitting}
\end{figure*}

In order to obtain the cluster potential and the scale radii for the two populations, we model the cumulative mass profile of the cluster. A mass density model does not give the probability (likelihood) that we will find a star of given mass at some radius, $P(m, r)$. Thus, given a set of (cluster-centric) radius and mass of $N$ stars, $\{r_i,\; m_i\}$, we consider a transformed dataset of cumulative mass in increasing order of radius. Let $j$ be the index that sorts $r_i$ in non-decreasing order, i.e., $r_j \le r_{j+1}$. Then, the transformed data is
\begin{equation}
    \{ r_j,\; M_j \} \equiv \{ r_j,\; \displaystyle\sum^j_{k=1} m_k \}
\end{equation}
and we fit the cumulative mass profile model to $\{ r_j,\; M_j \}$. Fig.~\ref{fig:cumulative-mass-fitting} shows the cumulative mass profile from the transformed data for the entire data (a), for high-mass stars (b; $r_c<10$~pc), and for low-mass stars (c; $r_c<10$~pc).

For the widely-used \citet{Pl11} model, the mass profile is 
\begin{equation}
    M(r) = M_0 \frac{ r^3 }{ (r^2 +a^2)^{3/2} }.
\end{equation}
We also consider an extension
\begin{equation}
    M(r) = M_0 \frac{ r^3 }{ (r^n +a^n)^{3/n} },
    \label{eq:generalized-plummer}
\end{equation}
which we refer to as the Generalized Plummer family.

\wyn{Of course, the enclosed mass $M(r)$ is well-defined even for aspherical density laws. However, if the matter distribution is spherically symmetric, then the density is}
\begin{equation}
    \rho(r) ={3M_0 a^n\over 4 \pi} {1\over (a^n + r^n)^{1+3/n}}
\end{equation}
All the models have a roughly constant density core, but
$\rho \rightarrow r^{-(3+n)}$ as $r \rightarrow \infty$.
Thus, the family is flexible enough to fit densities that decay less steeply than the $r^{-5}$ fall-off of the Plummer model, which is the case for the tidal tails of the Hyades. Note that, in comparing between different models in this family, it is fairer to use the half-mass radius which is
\begin{equation}
\label{eq:rhalf}
    \rh = a \left( 2^{n/3}-1\right)^{-1/n},
\end{equation}
rather than the scalelength $a$.

We fit the data using Pystan, a python interface to the statistical modelling platform Stan \citep{Carpenter2017-stan}. In all fitting, we assume a 10~\% Gaussian noise on the stellar mass of individual sources, which determines the Gaussian noise $\sigma_j$ for each $M_j$ via the sum of variances. The likelihood is, then, 
\begin{equation}
    M_j \sim \mathcal{N}(M(r_j),\,\sigma_j)
\end{equation}
where $\mathcal{N}(\mu,\,\sigma)$ is the 1D Gaussian distribution with the mean $\mu$ and standard deviation $\sigma$. We use No-U-Turn Sampler to sample the posterior distribution of model parameters, and check the Gelman-Rubin statistic $\hat R$ of the posterior draws to ensure that the chains have converged \citep{gelman1992-rhat,vehtari2019-rhat}. We assumed (improper) uniform priors on the parameters. Because the data are highly informative, the choice of prior distribution has little effect on the inference.
\so{In this model, we ignore the parallax uncertainties. The median parallax signal-to-noise ratio of the sample is $257$,
which corresponds to a distance error of $0.2$~pc. This is much smaller than the typical scale and the scale radius difference of a few pc.}

We first fit the entire data (cluster and tails) with a two-component Generalized Plummer model, as reported in the upper panel of  Table~\ref{tab:fit-result}. We find that at least two distinct components are needed to adequately describe the cumulative mass profile over the entire radius range populated by cluster and tails. In particular, a single Generalized Plummer profile is a poor fit to the entirety of the data. The best-fit model and its components are compared to the data in Fig.~\ref{fig:cumulative-mass-fitting} (a). The inner component is Plummer-like ($n = 1.822$) with a scale radius of $a = 4.031$~pc, corresponding to a half-mass radius $\rh = 5.751$~pc. This is larger than found in earlier studies. For example, using a pure Plummer model and shallower data, \citet{Pe98} found $\rh \approx 3.8$ pc, whilst \citet{Ro11} preferred $\rh \approx 4.0$ pc. On the other hand, the outer component prefers smaller $n$ ($= 1.360$), indicating that the density of tails falls off less steeply than the Plummer, namely $\rho\sim r^{-4.36}$.
% SO: where is the conversion from volume to line density? sedate compared to what?
This though is a little misleading, as the tails are not spherically distributed about the cluster centre. If we convert this three dimensional density to a line density along the tails, then $\rho \sim d^{-2.36}$ where $d$ is length along the tail. 

As can be seen in Fig.~\ref{fig:cumulative-mass-fitting} (a), the inner component dominates within $\lesssim 20$~pc with little contribution from the outer component.  Given that the inner component in the best-fit two-component model is Plummer-like ($n\approx 2$), we fit the normal Plummer model to the cumulative mass profiles of the high-mass and low-mass subsamples, presenting the results in the lower panel of Table~\ref{tab:fit-result}. Fig.s~\ref{fig:cumulative-mass-fitting} (b) and (c) show both fits, together with the data. The Plummer scale radii for high-mass and low-mass stars are $3.741$~pc and $6.208$~pc, respectively. Using eq.~(\ref{eq:rhalf}), these correspond to half-mass radii of $4.881$~pc and $8.100$~pc.

The central parts of the Hyades are significantly depleted in low mass stars. Qualitative evidence that the Hyades is mass segregated has of course been presented before~\citep[e.g.,][]{Pe98,Go13}, but we have here provided a quantitative demonstration that the low-mass population is much more spatially extended than the high mass.

\begin{table}
    \centering
    \begin{tabular}{c|rr|rr}
        \hline
        \multicolumn{5}{c}{all stars (two-component Generalized Plummer model)} \\
        \hline
              & \multicolumn{2}{c}{component 1} & \multicolumn{2}{c}{component 2} \\
        \hline
        $a$ [pc]  & 4.031 & 0.007 & 33.33 & 0.73 \\
        $\rh$ [pc]  & 5.751 &  & 69.35 &  \\
        $M_0$ [$M_\odot$] & 313.3 & 1.3   & 261.8 & 3.5 \\
        $n$   & 1.822 & 0.009 & 1.360 & 0.024 \\
        \hline\hline
        \multicolumn{5}{c}{$r_c<10$~pc (Plummer model)} \\
        \hline
              & \multicolumn{2}{c}{high mass} & \multicolumn{2}{c}{low mass} \\
        \hline
        $a$ [pc]  & 3.741 & 0.009 & 6.208 & 0.010\\
        $r_{\rm h}$ [pc] & 4.881 & &8.100 &\\
        $M_0$[$M_\odot$] & 207.5 & 0.56 & 117.3 & 0.28\\
        \hline
    \end{tabular}
    \caption{
        Best-fit parameters of cumulative mass profile.
        The parameters $a$, $M_0$, and $n$ are from the Generalized Plummer model        (eqn~\ref{eq:generalized-plummer}) while $r_h$ is the calculated half-mass radius.
        }
    \label{tab:fit-result}
\end{table}

% NOTE: Multiplicative factor to scale radius to get half-mass radius
% for generalized Plummer model
% n =  0.5, R_h/a =   66.680
% n =  0.7, R_h/a =   12.007
% n =  1.0, R_h/a =    3.847
% n =  1.5, R_h/a =    1.800
% n =  2.0, R_h/a =    1.305    <-- plain Plummer.
% n =  3.0, R_h/a =    1.000
% n =  4.0, R_h/a =    0.901

\begin{figure}
    \centering
    \includegraphics[width=0.95\linewidth]{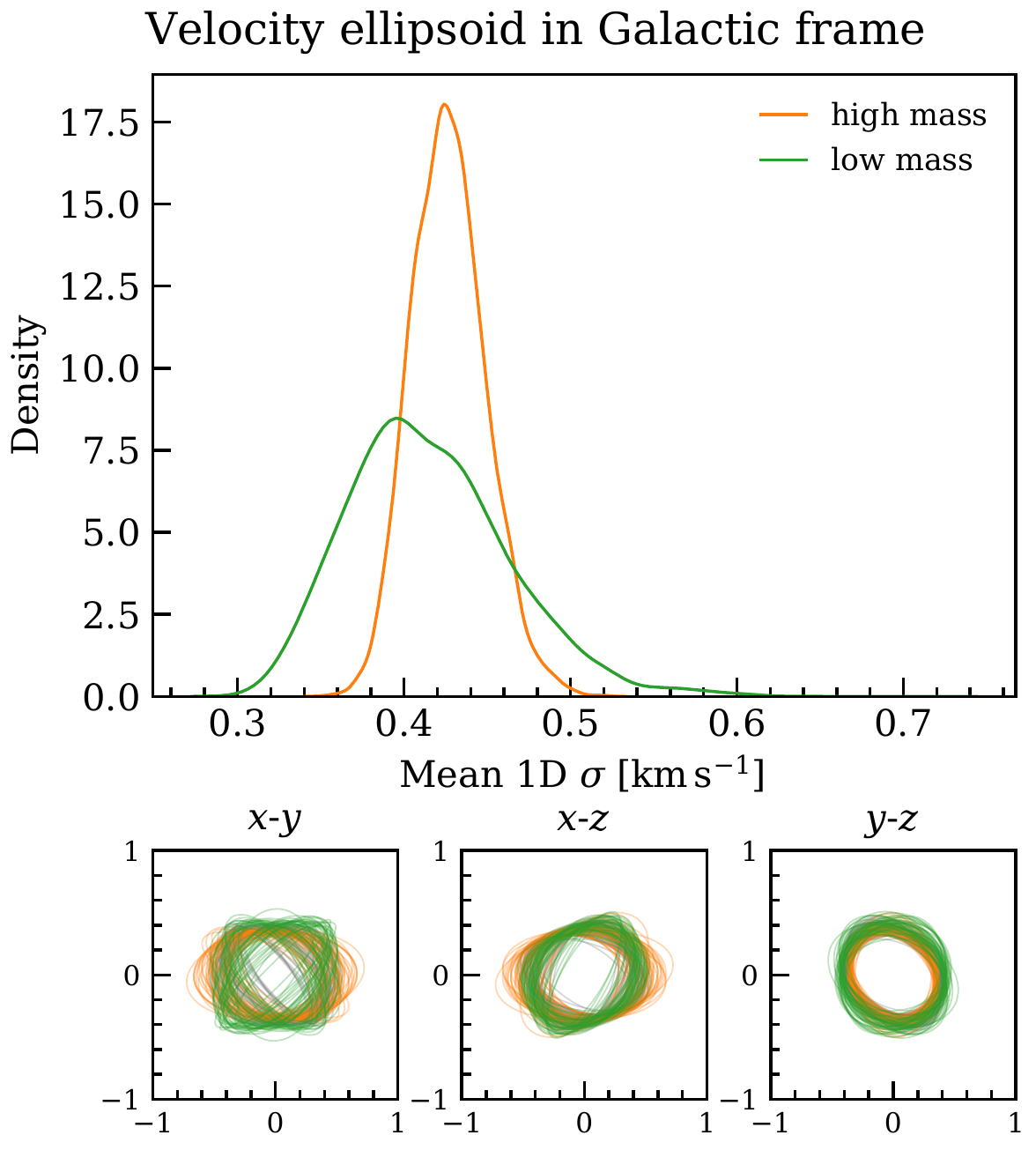}
    \caption{Comparing the average 1D velocity dispersion of high-mass and low-mass stars. Top: The posterior distribution of the 1D velocity dispersion for the high mass and low mass populations within the tidal radius of the Hyades. Bottom: Random draws from the posterior distributions of the velocity ellipsoid for the high mass and low mass populations, projected in the ($x,y,z)$ principle planes. The orthogonal ($x,y,z$) coordinate system is Sun-centered, with positive $x$ in the direction towards the Galactic centre and positive $y$ in the direction of Galactic rotation.}
    \label{fig:kinematics}
\end{figure}
\begin{figure}
    \centering
    \includegraphics[width=0.95\linewidth]{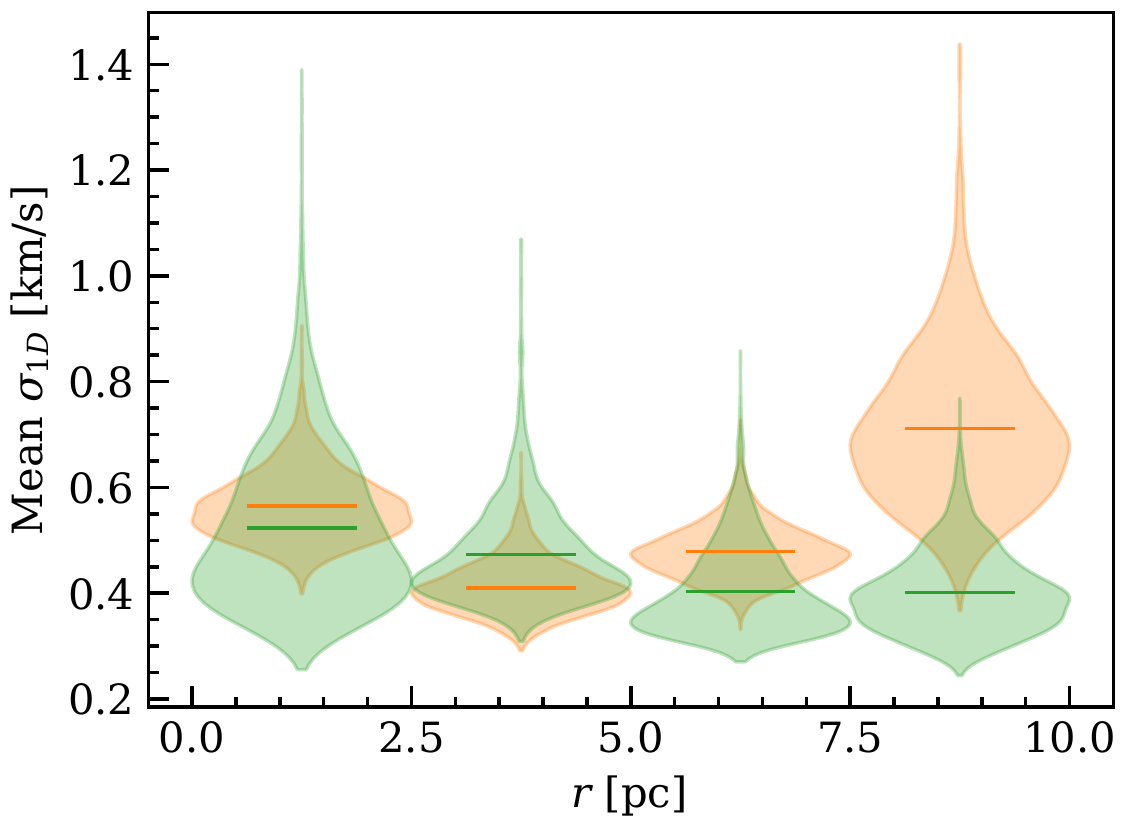}
    \caption{Posterior probability density distribution of the mean 1D velocity dispersion for subsamples binned in $r$ for the high (orange) and low (green) mass populations. For each violin, the maximum width is set to the bin size. The relative width at given $y$-axis value is proportional to the posterior probability density. Horizontal bars indicate the mean of the posterior distribution.}
    \label{fig:rc-sigma1d}
\end{figure}

\subsection{Kinematics}

\begin{table}
    \centering
    \begin{tabular}{lrrrr}
    \hline
     Parameter & \multicolumn{2}{c}{high mass} & \multicolumn{2}{c}{low mass}\\
    \hline
    $v_x$                     &   -6.117 & (  -6.192,    -6.036)&  -6.040 & (  -6.150,    -5.916)\\
    $v_y$                     &   45.626 & (  45.524,    45.739)&  45.717 & (  45.457,    45.983)\\
    $v_z$                     &    5.522 & (   5.456,     5.592)&   5.569 & (   5.471,     5.670)\\
    mean $\sigma_\mathrm{1D}$ &    0.427 & (   0.386,     0.467)&   0.415 & (   0.329,     0.502)\\
    \hline
    \end{tabular}
    \caption{
        Summary of the kinematics for the high-mass and low-mass population.
        For each parameter, we quote the mean of the posterior samples and the 94\% highest posterior density credible interval. Mean velocities are in the ICRS coordinate frame.
        All values are in \kms.
    }
    \label{tab:kinematics}
\end{table}

We model the kinematics of high-mass and low-mass stars separately using the forward-modelling method for internal motions of stars in a cluster presented in OE. The velocity field is described by a linear velocity gradient describing rotation and shear, together with an anisotropic velocity dispersion matrix. OE applied their method to stars in both the cluster proper and the tidal tails, allowing for contaminants from background populations and missing radial velocities for some of the stars. They found that the cluster is isotropic with marginal evidence for expansion, but no rotation. However, OE did not examine the kinematics of the cluster stars as a function of their mass.

Henceforth, we use only the velocities of stars within the tidal radius and examine the global kinematical quantities as a function of mass. We find that there is no velocity gradient (in other words, no evidence for rotation and/or shear) for the high-mass and low-mass populations separately within the tidal radius. The top panel of Fig.~\ref{fig:kinematics} compares the mean one-dimensional velocity dispersion $\sigma = \sqrt{(\sigma_x^2 + \sigma_y^2 + \sigma_z^2)/3}$
of the two populations, which are consistent with each other ($\sigma \approx 0.42$~\kms). The mean velocity of the two populations when independently modeled are the same, indicating that there is no systemic relative motion between the two. The summary statistics of the posterior distributions for the mean velocity and mean 1D velocity dispersion are presented in Table~\ref{tab:kinematics}.

The bottom row of Fig.~\ref{fig:kinematics} shows 50 random posterior draws of the velocity ellipsoid in the Galactic frame for high-mass (orange) and low-mass (green) populations. \wyn{While there is a hint that the velocity ellipsoid for the high-mass population is more elongated in the Galactic radial direction ($\sigma_x$) compared to the low-mass population,} we do not find any significant evidence for a difference between the two samples. 

Finally, we can study the variation of the velocity dispersion with radius for the high and low mass populations separately, as shown in Fig.~\ref{fig:rc-sigma1d}. The relative width of the violins shows the posterior probability density of the mean 1D velocity dispersion, with the horizontal bar denoting the mean. Aside for the outermost bin of the high mass population -- which contains few stars and is therefore not wholly trustworthy -- there is no significant difference in the spatial variation of the velocity dispersion of the two populations in the Hyades. The dispersions are slightly larger in the centre, decrease, but then increase again at the outskirts. The dominant feature is that there is not much variation with radius. To within the uncertainties, both populations have nearly uniform velocity dispersions and there is no variation in the degree of equipartition with cluster-centric radius.

\wyn{\citet{Bi16} provide a fitting formula that describes the mass dependence of the velocity dispersion of partially relaxed populations in globular clusters. It is deduced from a set of globular cluster simulations, so its applicability to open clusters is unclear. Even so, the formula does predict that the high-mass and low-mass populations in the Hyades will have similar velocity dispersions. Specifically, the Hyades has experienced $\approx 10$ relaxation times, so the kinematic difference between the two stellar mass populations is $\lesssim 10\%$. This is the same order of magnitude as the uncertainty in dispersion reached with the Gaia DR2 data (see Table~\ref{tab:kinematics}) and so the difference is presently unmeasurable. Note that the Gaia early Data Release 3 does not improve matters, as the velocity dispersions are limited by the sampling noise rather than the measurement precision.}
 
We conclude that the Hyades is far from a state of complete energy equipartition. This would imply that the velocity dispersion scales with stellar mass $m$ as $\sigma \propto m^{-1/2}$. This is not what is observed. 

\subsection{Effects of Binaries}
\label{sub:effects-of-binaries}

Binaries play a fundamental role in the dynamics of star clusters. The division between ‘hard’ and ‘soft’ binaries is the separation at which the orbital velocity is equal to the mean 1d velocity dispersion of the cluster. For a system mass of $1 M_\odot$, then the Hyades hard binaries have separation $\lesssim 5000$ AU and a period $\lesssim 200$ yr. If not properly accounted for, binaries may also bias the observationally determined properties of clusters. In this Section, we investigate the relative binarities of the low-mass and high-mass population, and assess their impact on the scale radii and velocity dispersions.

\subsubsection{Comparison of binarity between high-mass and low-mass samples}

Despite the importance of binaries in understanding the dynamical evolution of star clusters and the pervasive manifestation of binarity in stellar observables, observational constraints remain sparse as often they require a dedicated survey using spectroscopy (e.g., WYIN Open Cluster Survey; \citet{wosc}) or adaptive optics \citep[e.g.,][]{bouvier2001-praesepe-ao}. Reconstructing binary population accounting for respective observational biases from the methods of detection is a complex problem \citep[e.g.,][and the references therein]{moe2017}. Eventually, future data releases of \gaia\ time-series astrometry may dramatically change this situation.

With regard to the \gaia\ data released so far, binaries may be detected through the offset to the single-star main sequence on the colour-magnitude space \citep[particularly assisted by the excellent precision of the \gaia\ parallaxes;][]{liu2019-smoking-gun,li2020-binary-cmd}, the goodness-of-fit to the single-source astrometric solution \citep{belokurov2020,penoyre2020}, the anomalously large radial velocity \lq\lq error" for sources in the RVS sample (Boubert et al. in prep), and the on-sky acceleration often called the proper motion anomaly \citep{brandt2018,kervella2019}.  Finally, for resolved wide binaries, we can look for close pairs of sources with proper motion difference below the orbital velocity threshold \citep{elbadry2018,deacon2020}.  Each of these detections of binarity are sensitive to different regions of the binary parameter space.

Utilizing all the \gaia\ data to provide the best constraints on the binary population of the Hyades is a substantial effort, beyond the scope of this paper. However, insights into the relative binary fraction between high-mass and low-mass samples can still be obtained through quantities readily available in \gaia\ DR2. 

The RUWE, or re-normalized unit weight error, of \gaia\ provides a goodness-of-fit statistic of the single-source astrometric fits for each source~\citep[e.g.,][]{Gaia-dr3-release}. Since the trajectory of an unresolved binary has extra variance due to orbital motion, a high RUWE value can indicate that the source is an unresolved binary~\citep[e.g.,][]{belokurov2020,penoyre2020}. Recently, Penoyre et al (in prep) showed that the cut RUWE $> 1.4$ extracts more than $85 \%$ of such unresolved binaries within 50 pc in the \gaia\ Universe Model Snapshot of~\citet{Ro12}. They conclude that for binaries within 50~pc and with periods $T$ between $10^{-2}\lesssim T \lesssim 10^2$ yrs, the RUWE statistic is a powerful diagnostic of unresolved companions. For the Hyades, we conclude that RUWE is a good probe of the hard binary population.

\begin{figure}
    \centering
        \includegraphics[width=0.95\linewidth]{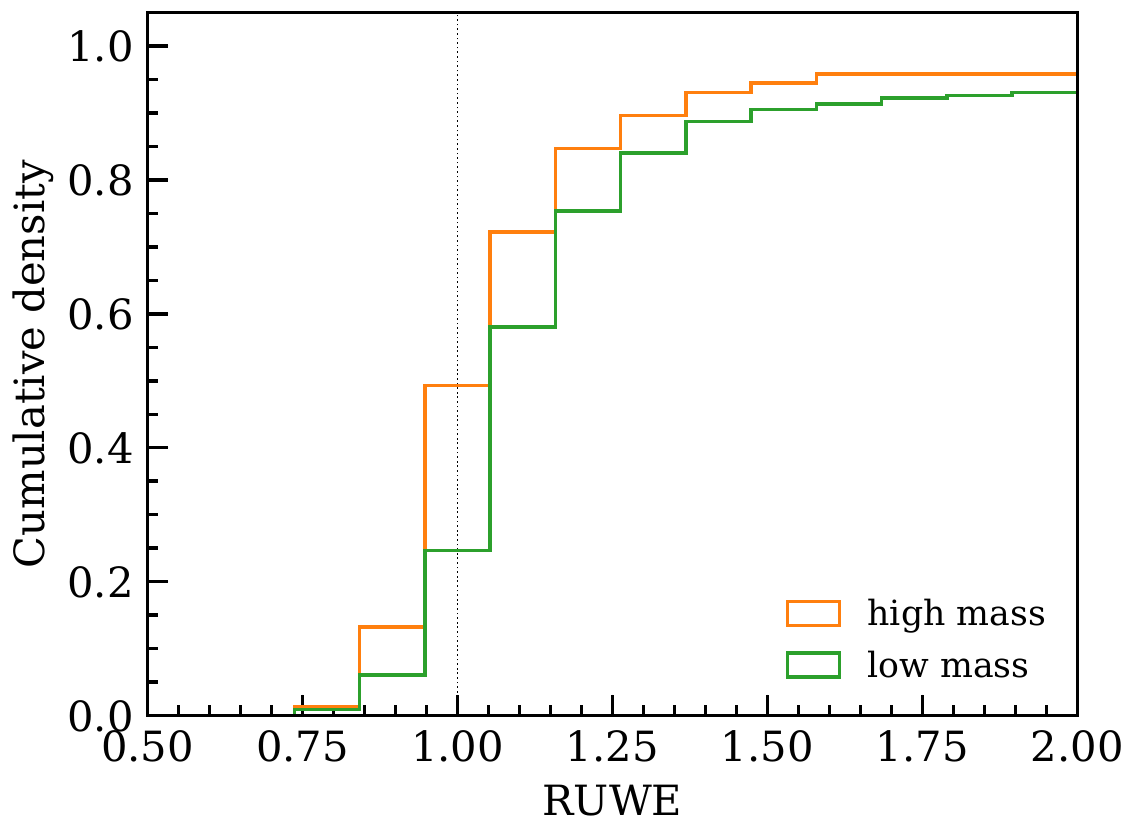} 
    \caption{
        RUWE distributions of the high-mass and low-mass samples.  RUWE is a goodness-of-fit statistic of the single-source astrometric fits available for all sources in \gaia\. A high value of RUWE may indicate that the source is an unresolved binary as binary orbital motion adds extra jitter to the astrometry. We find that the RUWE distribution of the low-mass sample is more heavy-tailed to larger values, indicating that low-mass stars may have higher unresolved binary fraction than the high-mass stars.
        }
    \label{fig:ruwe-dist}
\end{figure}

We compare the RUWE distribution of the the low-mass and high-mass samples in Fig.~\ref{fig:ruwe-dist}. We find that the RUWE distribution of the low-mass population is more heavy-tailed to large values than the high-mass sample. At face value, this indicates that the binary fraction is larger for the more dispersed low-mass population than the centrally concentrated high-mass population in the Hyades. Adopting a RUWE threshold of 1.4, $15$\% ($N=35$) and $8.3$\% ($N=12$) 
would be classified as astrometric binaries for the low-mass and high-mass samples, respectively. The RV binary flag from Boubert et al. (in prep), which labels sources showing RV time variability above the expected level from single stars, also does not suggest a high binary fraction. Of 93 high-mass stars ($65$\%) that are in the \gaia\ DR2 RVS sample, only 5 sources show significant RV variability that is evidence of binarity. Though our results on binarity remain tentative, we see no evidence of an excess of hard binaries in the core. By contrast, in the open cluster simulations of \citet{Sp16}, the formation of hard binaries in the centre is one of the mechanisms preventing core collapse. Certainly, models of the Hyades in which a high proportion of stars in the central few parsecs are binaries~\citep[e.g.,][]{Kr95} are disfavoured by the existing data.
We note that a deficiency of binaries in the inner region of another intermediate age ($\approx 300$ Myr) open cluster, NGC 3532, was also recently reported in \citet{li2020-binary-cmd} by modelling unresolved binaries in the \gaia\ color-magnitude diagram.

\subsubsection{Effect on scale radius}
\label{sub:effect-on-scale-radius}

\begin{figure*}
    \centering
    \includegraphics[width=0.65\linewidth]{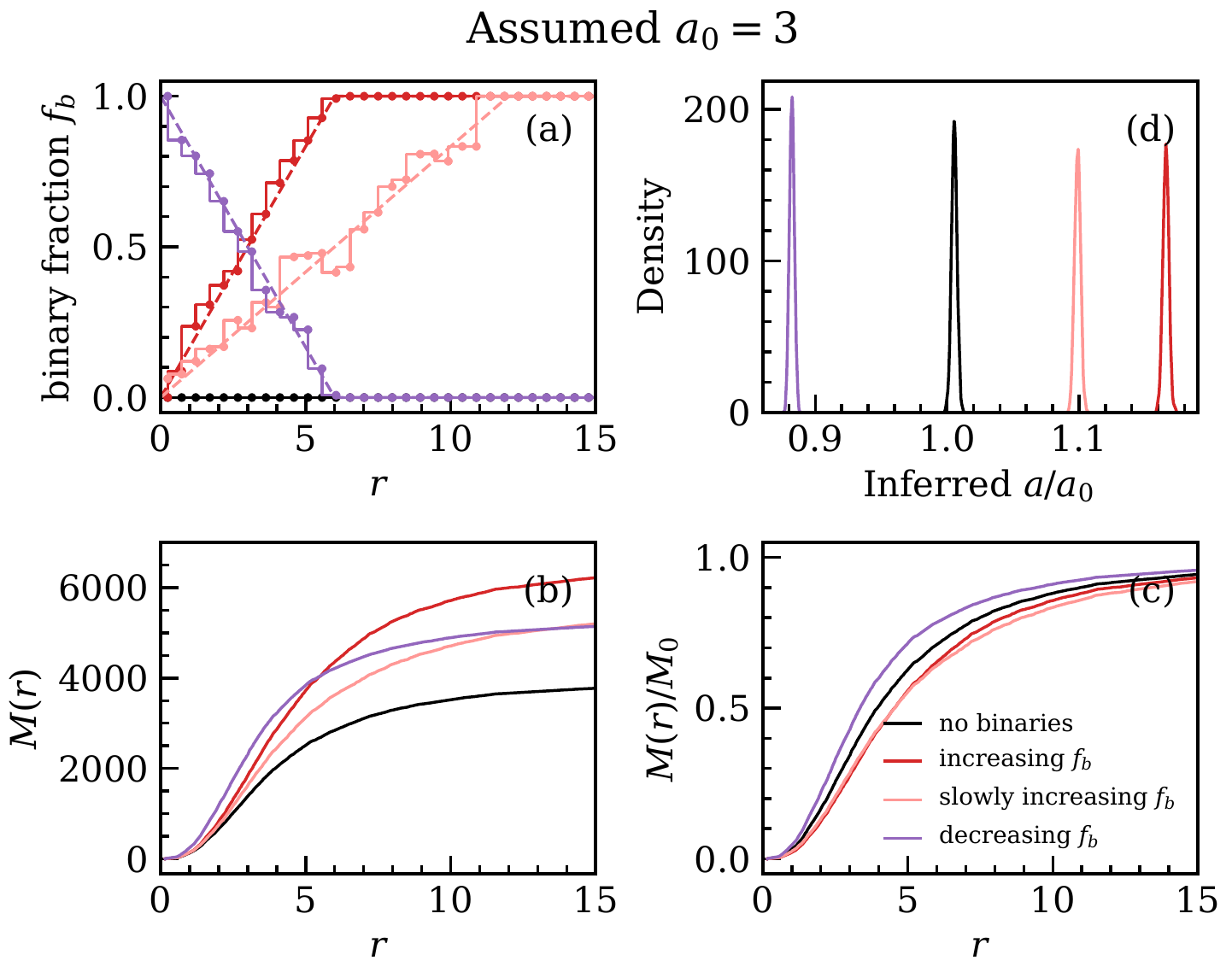}
    \caption{
        Systematic bias on the inferred Plummer scale radius $a$
        due to radially-changing binary fraction when companion mass
        is ignored (i.e., all stars are treated as single stars).
        In this experiment described in Section~\ref{sub:effect-on-scale-radius},
        all stars have mass $1$ (arbitrary unit) and all binaries are equal-mass binaries.
        Panel (a) shows differing scenarios of radially-changing binary fraction.
        Panel (b) and (c) shows the cumulative mass profiles and cumulative mass fraction
        for each case.
        Finally, panel (d) shows the posterior distributions of the inferred scale radius in each case
        when mass of the companions are correctly included ($a$) compared to when all stars
        are treated as single stars ($a_0$).
        The systematic bias in these rather extreme cases of radially-changing binary fraction
        is $\approx 10-16$\%.
    }
    \label{fig:binfrac-experiment}
\end{figure*}

As mentioned in Section~\ref{sub:mass-assignment}, our mass assignment method does not take binarity into account, which could result in assigning a stellar mass to a source off by at most a factor of $\approx 2$. While a better modelling of the colour-magnitude diagram is out of scope for this work, we can consider the effect of this on constraining the Plummer scale radius. For the scale radius, what matters is how fast or slow the enclosed mass fraction is rising with radius. Thus, the scale radius is only affected when the binarity changes as a function of radius.

We investigated the effect with simplified simulated data of a cluster of $4000$ stars. Stars have mass of $1$ in the case of single and $2$ in the case of binary. The radii were sampled from the Plummer distribution function assuming a scale radius of $a_0=3$. We considered four different scenarios of radially changing binary fraction, $f_b$: no binaries, binary fraction increasing from $0$ to $1$ in one scale radius, binary fraction increasing from $0$ to $1$ in two scale radii, and binary fraction decreasing from $1$ to $0$ in one scale radius. The top left panel (a) of Fig.~\ref{fig:binfrac-experiment}
shows the functional form of $f_b(r)$ for the four cases as dashed lines of different colours, with corresponding simulated data shown as histograms. The resulting cumulative mass profile $M(r)$ and normalized cumulative mass profile $M(r)/M_0$ are shown in the bottom left (b) and bottom right (c) panel, respectively. We fitted each simulated cumulative mass profile the same way as we have done for the actual data as described in Section~\ref{sub:cumulative-mass-profiles}. The resulting posterior probability density distribution of the scale radius is shown in the top right panel of Fig.~\ref{fig:binfrac-experiment}. As expected, if there are no binaries, we recover the assumed scale radius from the cumulative mass profile. However, if there are more binaries at the center, the inferred scale radius is smaller than the case with no binaries as the fractional enclosed mass profile rises faster. The opposite is true for the increasing binary fraction. The scale radius changes by $\approx 10-16$\%. Since we do not expect the binary fraction to change so dramatically nor for all binaries to be equal-mass resulting in maximally-incorrect assigned mass in reality, we consider this to be the limiting systematic bias on our scale radii due to radially-changing binarity. Thus, we conclude that the difference in scale radii between the low-mass and high-mass populations is still significant beyond this systematic bias.

\subsubsection{Effect on kinematics}

When measuring the internal dispersion of a cluster from its kinematic snapshot either from radial velocities alone or in combination with proper motions, unknown binary population may inflate the dispersion by adding extra jitter due to their orbital motion \citep{mcconnachie2010,Ko11,minor2019}.  Similar to OE, we tested whether the posterior distributions of velocity dispersion for high-mass and low-mass samples change when we exclude sources with high RUWE (RUWE $>1.4$) and flagged as showing significant RV variability (Boubert et al. in prep). We found no significant difference in the posterior distributions, and conclude that binaries are not inflating the velocity dispersion of either samples.

\section{The Equilibria of Mass-Segregated Populations}
\label{sec:models}

We begin with a derivation of the conditions under which energy equipartition holds. Next, we derive a new instability of mass-segregating systems by explicitly taking into account a finite escape speed, before applying our results to the Hyades cluster.

\subsection{Energy Equipartition}
\label{sec:thermal}

We study the thermal equilibrium of two populations composed of stars with mean masses $m_1$ and $m_2$. Without loss of generality, we take $m_1 > m_2$, so population 1 is composed of stars with higher mass. Under gradual mass segregation, the higher mass stars slowly sink and so population 1 becomes more centrally concentrated than population 2, so we expect $r_1 < r_2$. Simultaneously, the velocity dispersion of population 1 becomes smaller, and of population 2 larger. 

The equipartition of energy states that
\begin{equation}
\label{eq:equipartition}
m_1 \sigma_1^2 = m_2 \sigma_2^2
\end{equation}
where $\sigma_1$ and $\sigma_2$ are the velocity dispersions of populations 1 and 2. Equipartition is often seen as the natural endpoint of the collisional evolution of multi-mass systems. Let us start by understanding the assumptions in the derivation of equipartition.

Consider a single object in the component with the high-mass stars. It loses kinetic energy due to gravitational encounters  with objects in the component with low mass stars. If the velocity of the single object is $v$, the rate of change of kinetic energy per unit mass is ~\citep[][eq. 7.90]{BT}
\begin{equation}
\label{eq:energyloss}
\left\langle 
\frac{\partial}{\partial t} \left( \frac{v^2}{2} \right) 
\right\rangle
= 
16\pi^2 G^2 \ln \Lambda 
\left( 
m_2 \int^{\infty}_v w F_2 dw -
m_1\int^v_0\frac{w^2}{v} F_2 dw 
\right),
\end{equation}
with $F_2$ as the distribution function (DF) of the low mass population. Here, $\langle \cdot \rangle$ denotes an ensemble average, whilst the Coulomb factor $\Lambda$ is the ratio of the maximum and minimum impact parameters, $b_{{\rm max}}$ and $b_{{\rm min}}$ ~\citep[][eq. 1.33b]{BT}. Note in eq~(\ref{eq:energyloss}) that only population 2 stars moving faster than $v$ contribute to the heating, whilst only those moving slower contribute to the cooling.

By multiplying equation (\ref{eq:energyloss}) by the DF of the high mass stars $F_1$ and integrating over the whole of velocity space, we obtain the rate of change of kinetic energy per unit volume of the stars in population 1: 
 \begin{equation}
\label{eq:ke}
{dK\over dt} =
4\pi \int_0^\infty  v^2  F_1
\left\langle 
\frac{\partial}{\partial t} \left( \frac{v^2}{2} \right) 
\right\rangle dv
 \end{equation}

Let us now assume that both the distributions $F_1$ and $F_2$ are isothermals, so that
\begin{equation}
\label{eq:isothermaldfs}
    F_i = {\rho_i \over (2\pi\sigma_i^2)^{3/2}}\exp \left(-{v^2\over 2 \sigma_i^2}\right), \qquad\qquad i=1,2
\end{equation}
with $\sigma_1$ and $\sigma_2$ constants. We now insert these choices into eq~(\ref{eq:ke}) and obtain
\begin{equation}
\label{eq:dkdt}
 {dK\over dt} = 4\sqrt{2\pi} {G^2\over \sigmaT^3} \left(m_2\sigma_2^2 - m_1 \sigma_1^2 \right) \rho_1\rho_2 \ln \Lambda,
\end{equation}
where $\sigmaT^2 = \sigma_1^2 + \sigma_2^2$. For equilibrium, that is no net energy flow between populations 1 and 2, then $dK/dt=0$ and so we obtain the familiar result of energy equipartition~(\ref{eq:equipartition}), namely
\begin{equation}
    {m_2\over m_1} = {\sigma_1^2\over \sigma_2^2}
    \label{eq:equipartone}
\end{equation}

This derivation makes clear that equipartition holds only under very strong assumptions. First, all real clusters have an escape speed, yet eq.~(\ref{eq:equipartone}) holds only when the velocity integrals are carried out on an infinite domain. The second velocity moments are always sensitive to the tails of the velocity distribution, and so the existence of an escape speed can change their values considerably. Secondly, if the densities $\rho_1, \rho_2$ and dispersions $\sigma_1$, $\sigma_2$ depend on position, then eq.~(\ref{eq:energyloss}) needs a further integration over volume. This complexity does not arise with isothermals, which of course have constant dispersions.

\citet{Sp69} realized that equipartition can be unachievable. He used an idealized cluster composed of two populations of stars with masses $m_1$ and $m_2$ (with $m_1>m_2$). He argued that they cannot reach equipartition if 
\begin{equation}
\label{eq:spitzer}
M_1 > 0.16\, M_2 \left( {m_2\over m_1} \right)^{3/2}
\end{equation}
where $M_1$ and $M_2$ are the total masses of populations 1 and 2, respectively. The Hyades cluster is unstable according this criterion, but it is unclear that Spitzer's analysis really applies to the Hyades. Spitzer's instability is actually derived under the assumption that the total mass in heavier objects is less than that in lighter ones $M_1 <M_2$, but the individual masses satisfy $m_1 \gg m_2$.  In  other words, a few very heavy objects decouple and contract to form a subsystem at the centre of the cluster.

Spitzer's instability is often cited as an important factor in the failure of clusters to reach equipartition, either observationally or in simulations~\citep[e.g.,][]{Tr13,Sp16}. However, its actual importance is unclear. Other physical processes may also play an important -- perhaps dominant -- role in thwarting equipartition.

\begin{figure}
    \centering
    \includegraphics[width=0.9\linewidth]{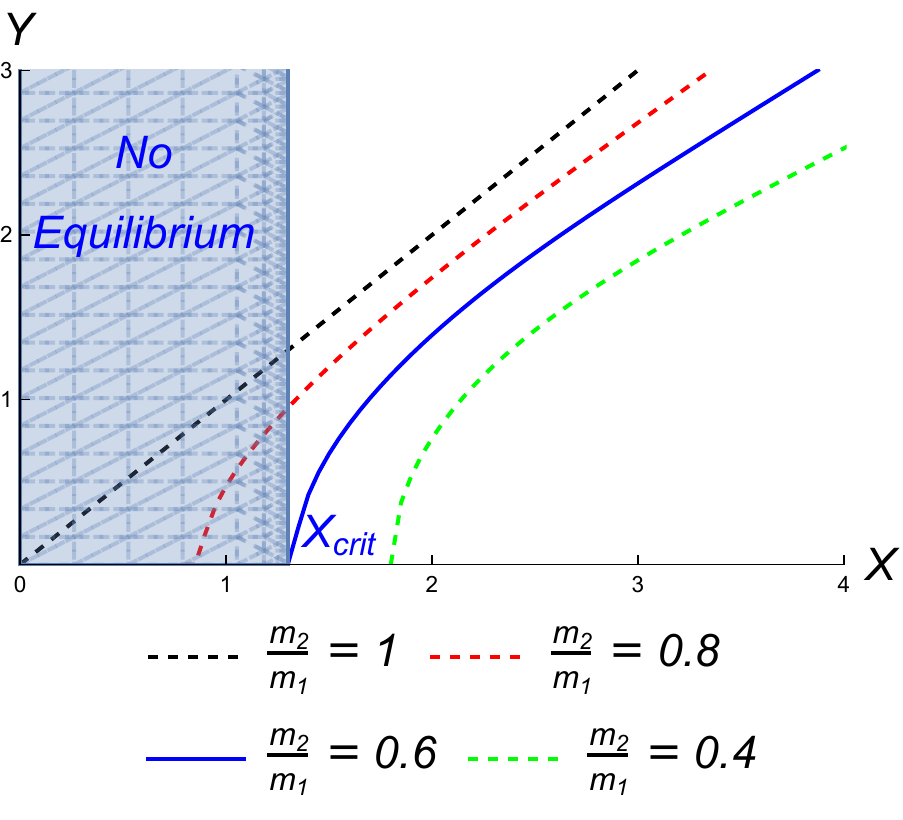}
    \includegraphics[width=0.9\linewidth]{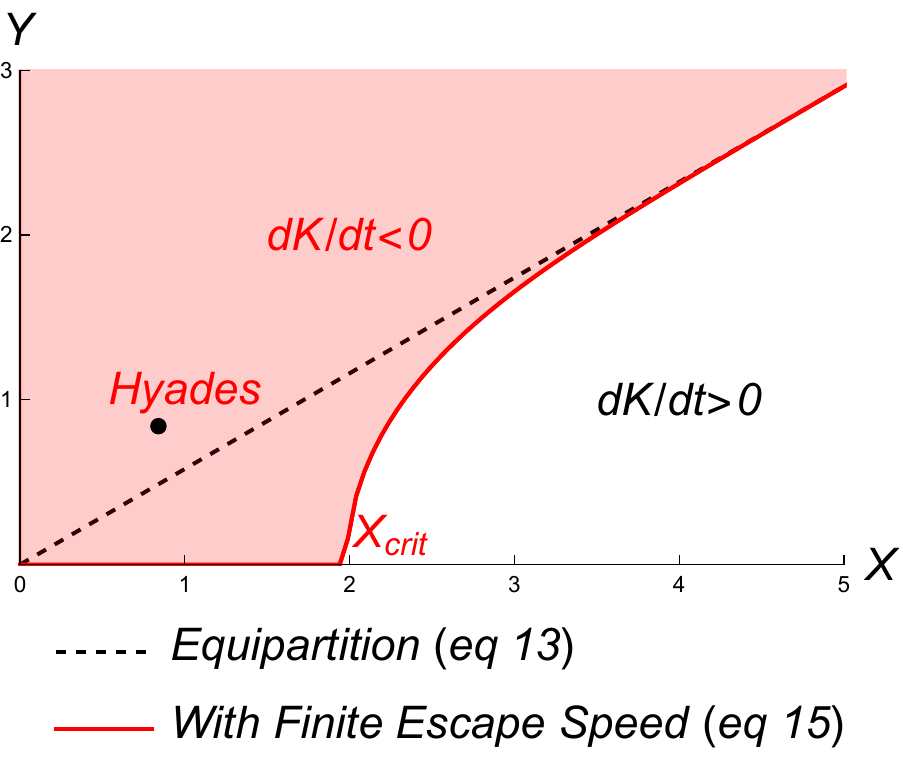} 
    \caption{Top: Thermal equilibria in the plane of $X = V/(\sqrt{2}\sigma_1), Y = V/(\sqrt{2}\sigma_2)$ for different mass ratio populations. Note the existence of a mass segregation instability. If $m_2/m_1 \neq 1$, there is always a regime $X<\Xcrit$ for which no equilibrium is possible. This is illustrated for the case $m_2/m_1 =0.6$ for which $\Xcrit = 1.33$ and no equilibrium exists if $X< 1.33$, or equivalently, $\sigma_2 > 0.53 V$.
    Bottom: Thermal equilibrium of the high and low-mass populations in the Hyades cluster ($m_2/m_1= 0.337$). The dashed line shows the equilibria corresponding to energy equipartition (eq~\ref{eq:equipartone}). The full red line shows the effects of a finite escape speed (eq~\ref{eq:genequipart}). The red shaded region, in which the location of the Hyades is marked, corresponds to systems in which there is no equilibrium but a persistent flow of energy out of the cluster $dK/dt <0$. }
    \label{fig:equi}
\end{figure}

\subsection{Effect of the Escape Speed}

Suppose we introduce an escape speed $V$. Once stars are accelerated through encounters to velocities greater than $V$, they are lost to the cluster. The rate of change of kinetic energy per unit volume of the stars in population 1 becomes
\begin{eqnarray}
{dK\over dt} &=& 4G^2 \rho_1\rho_2\log \Lambda \left[ {2V(m_2\sigma_1^2 - m_1\sigma_2^2) \over \sigma_1\sigma_2\sigma_T^2}e^{-X^2-Y^2} \right.\nonumber \\
&+&\left. \sqrt{2\pi}\left( {m_1\over\sigma_1}e^{-X^2}\erf(Y) -{m_2\over\sigma_2} e^{-Y^2}\erf(X)\right.\right.\nonumber\\
&+& \left.\left. {m_2\sigma_2^2-m_1\sigma_1^2\over \sigmaT^3} \erf(\sqrt{X^2+Y^2})\right)\right]
\label{eq:kecomp}
\end{eqnarray}
where $X = V/(\sqrt{2}\sigma_1)$ and $Y = V/(\sqrt{2}\sigma_2)$. If $V\rightarrow \infty$, then we recover eq~(\ref{eq:dkdt}). 

Setting $dK/dt=0$ gives us
\begin{equation}
\label{eq:genequipart}
{m_2\over m_1} = {F(X,Y)\over F(Y,X)}.
\end{equation}
If the velocity dispersions of the two populations are equal, then $X = Y$ and so $m_1 = m_2$, consistent with equipartition. However, in general, matters are more complicated, as
\begin{eqnarray}
F(X,Y) &=&\sqrt{\pi} e^{Y^2}\Bigl[X(X^2+Y^2)\erf(Y)\nonumber\\ &-&{XY^3e^{X^2}\over\sqrt{X^2+Y^2}}\erf(\sqrt{X^2+Y^2})\Bigr] -2X^3Y.
\end{eqnarray}
So, the result deviates from equipartition, especially if the velocity dispersions are comparable to the escape speed.

% WIP: the command \wyn does not work well with equations inside.

Curves of eq~(\ref{eq:genequipart}) for different mass ratios are  shown in the upper panel of Fig.~\ref{fig:equi}. They reveal the existence of a new mass segregation instability. Let $\Xcrit$ be the intersection of any curve with $Y=0$, found by solution of
\begin{equation}
{6 \sqrt{\pi}\erf \Xcrit\,e^{\Xcrit^2} -4\Xcrit(2+2\Xcrit^2)\over  18\Xcrit + 3\sqrt{\pi} (2\Xcrit^2-3)\erf\Xcrit\, e^{\Xcrit^2}} ={m_2\over m_1}.
\end{equation}
Once $X<\Xcrit$, no thermal equilibrium is ever possible, as the velocity dispersion of the low mass population becomes formally infinite or the escape speed becomes vanishingly small. In practice, the cluster has disrupted, and all the low  mass stars have escaped.

For astrophysical applications, the mass ratio of stellar populations rarely exceeds a factor of 10, so $\Xcrit \lesssim 2$, or
\begin{equation}
\label{eq:instability}
    V/\sigma \lesssim 2\sqrt{2} 
\end{equation}
In other words, once the escape velocity is comparable to about three times the velocity dispersion, no thermal equilibrium is ever possible. This regime encompasses all the Galactic open clusters. \citet{Gn02} show that the Galactic globular clusters satisfy
\begin{equation}
    V/\sigma \approx 3.7 - (c-0.4)
\end{equation}
where $c$ is the concentration of the King model fit ($c =0.4-2.8$). So, many of the Galactic globular clusters are menaced by this instability as well. Both simulation and observational evidence against complete equipartition in globular clusters has been steadily mounting~\citep[e.g.,][]{Tr13,Pa16,Wa20}.

The physical origin of this instability is simple to understand. As encounters excite the velocities of low mass stars to greater than $V$, they are removed from the system. The low mass population continues to be heated by gravitational encounters with the high mass population, but there are fewer and fewer low mass stars available to absorb the heat. And when they do, the low mass stars are often removed from the cluster, as the velocity dispersion is comparable to the escape speed.

We note that this instability depends on the ratio of escape velocity to velocity dispersion. It is therefore distinct from Spitzer's instability, which depends on the total mass in the two populations. Of course, a dying cluster may be afflicted by both instabilities at the same time.

\subsection{The Status of the Hyades}

For the Hyades, the escape speed at the tidal radius can be estimated given a model for the cluster potential via $V = \sqrt{2\Psi(r_{\rm t})}$ with $r_{\rm t} =10$ pc. From Section~\ref{sec:Analysis}, the total density is well-modelled by the two Plummer profiles of the high and low mass populations. So, the cluster (relative) potential is
\begin{equation}
\Psi = {GM_1\over r_1} {1\over (1 + r^2/r_1^2)^{1/2}} +
{GM_1\over r_2} {1\over (1 + r^2/r_2^2)^{1/2}},
\end{equation}
where $M_1 = 207.5 M_\odot$ and $r_1 = 3.741$ pc, whilst $M_2 = 117.3 M_\odot$ and $r_2 = 6.208$ pc (from the lower panel of Table~\ref{tab:fit-result}). This gives $V$ as $0.50$ km s$^{-1}$, so that $X\approx Y \approx 0.84$ (using Table~\ref{tab:kinematics}). The escape speed is barely larger than the velocity dispersion. So, the Hyades lies in the regime in which the mass segregation instability in eq~(\ref{eq:instability}) applies. This is illustrated in the lower panel of Fig.~\ref{fig:equi}, for which the thermal equilibrium curve corresponding to the mass ratio $m_2/m_1 = 0.32/0.95 = 0.337$ is shown in red. The value of $\Xcrit = 1.98$ and so no equilibrium is even possible. The Hyades is located in the red shaded region for which $dK/dt <0$. This means that the Hyades must possess an energy flow through the cluster to maintain its structure.

We can evaluate this flow using the density and velocity dispersions. We have established that the two populations are well approximated by~\citet{Pl11} profiles 
\begin{equation}
    \rho_i(r) = {3M_i \over 4 \pi r_i^3 }{ 1\over (1 + r^2/r_i^2)^{5/2}}.
\end{equation}
with total masses $M_i$ and scalelengths $r_i$ given in Table~\ref{tab:fit-result}.

The velocity dispersion of each population is given by the Jeans equations. This is straightforward to solve, but it requires us to know the value of the velocity dispersion at the tidal radius as a boundary condition, namely:
\begin{equation}
    \rho_i \sigma_i^2 = \rho_i(r_{\rm t})\sigma^2_i(r_{\rm t}) -\int_r^{r_{\rm t}} dr\, \rho_i {d \Psi\over dr}
\end{equation}
Integrations with plausible choices of the boundary conditions show that the velocity dispersion profile has a mild dip roughly midway between centre and tidal radius, but is otherwise flattish. The profiles are consistent with the empirical results of Fig.~\ref{fig:rc-sigma1d}. So, we assume constant velocity dispersions with values taken from Table.~\ref{tab:kinematics} in what follows.

By integrating equation (\ref{eq:kecomp}) over the entire volume, we obtain the rate of change of total energy $E$ of the stars in population 1:
\begin{equation}
{\cal L}(r_1,r_2,m_1,m_2,X,Y) =\frac{dE}{dt} =\int d^3x {dK\over dt}.
\end{equation}
To evaluate this, we need to estimate the Coulomb logarithm term as
\begin{equation}
   \log \Lambda \approx \log {r_{\rm t} \sigma_1^2\over G (m_1+m_2)/2} \approx 6.5.
\end{equation}
This gives us the energy flow through the cluster as a function of the spatial scalelengths $r_!$, $r_2$ and average stellar masses $m_1, m_2$ of the two populations. 
Using our standard numbers for the Hyades, the energy flow is
\begin{equation}
{\cal L} \approx -9.5 \times 10^{-4} M_\odot\,{\rm km^2\, s^{-2} Myr^{-1}} 
\end{equation}
This flow is needed to maintain the current configuration of the low and high mass populations of the Hyades in a steady state. In fact, if the Hyades is disrupting, we  expect the flow to be still larger. 

OE estimated the current mass loss rate from the Hyades of $0.26 M_\odot {\rm Myr}^{-1}$. This by itself implies a flow of
\begin{equation}
{\cal L} \approx -2.1 \times 10^{-2} M_\odot\,{\rm km^2\, s^{-2} Myr^{-1}} 
\end{equation}
which is roughly a factor of twenty times larger. 

The mass loss in the Hyades is caused by two effects. The first is driven by tidal stripping in the Galactic tidal field. The second is evaporation of (predominantly) low mass stars from the cluster.

\section{Conclusions}

Open clusters are born in supersonically-turbulent Giant Molecular Clouds as bound systems of roughly a few hundred to a few thousand stars. At death, perhaps just tens of stars remain, most of them binary and multiple, until they too dissolve into the Galactic disc. Between birth and death, the life cycle of open clusters has to be pieced together from the clues around us -- the structure, stellar members and kinematics of the nearby open clusters. The proximity of the Hyades allows its stellar content to be scrutinised in great detail, especially with the advent of the \gaia\ satellite. It is a touchstone for our understanding of the evolution of open clusters.

The structure of the Hyades is heavily mass-segregated. The stars within the tidal radius of 10 pc have a Plummer-like profile with half-mass radius $r_{\rm h}$ of 5.75 pc. There are no very high mass stars in the Hyades now. As a convenient divider between the high and low mass populations, we use the \gaia\ colour BP-RP. The high mass population is defined by BP-RP $<2$ and has a mean mass of $0.95 M_\odot$. These  stars reside in a Plummer profile with a half-mass radius $r_{\rm h}$ of 4.88 pc. In contrast, the low-mass population (BP-RP $>2$ with mean mass $0.32 M_\odot$) has $r_{\rm h} = 8.10$ pc. Despite the differences in spatial extent, the kinematics of the high and low mass populations are very similar. They have isotropic velocity ellipsoids with mean 1d velocity dispersions $\sigma$ of 0.427 and 0.415 km\,s$^{-1}$ respectively.

At first sight, this is surprising. In Galactic dwarf spheroidals with multiple populations, the more extended population almost always has a larger velocity dispersion~\citep[e.g.,][]{Am12,Ko16}. In fact, for populations in equilibrium in the same gravitational field, a scaling between isotropic velocity dispersion and half mass radius must exist~\cite[e.g.,][]{Wa09,Wo10,Ag14}. However, the feeble Hyades is not in a state of equilibrium, but in a state of disintegration, driven by the mass loss from Galactic stripping and evaporation. 

The populations in the Hyades are not in energy equipartition ($\sigma\propto m^{-1/2}$). We have derived the condition for thermal equilibrium of populations in clusters with a finite escape speed. This led us to identify the criterion for a new mass segregation instability. {\it No thermal equilibrium is possible for populations with $V/\sigma \lesssim 2\sqrt{2}$.} This is distinct from the \citet{Sp69} instability, which holds true for an infinite escape velocity. In fact, the Hyades escape speed of $V \approx 0.50$ km is barely larger than the velocity dispersion, so it squarely falls within the regime of our new instability -- as do many other Galactic open and globular clusters). The weight of observational and simulation evidence against energy equipartition in globular clusters is now substantial~\citep[cf][]{Sp69,Tr13,Sp16}, but a full theoretical understanding remains wanting.

To maintain the two populations in the Hyades in their current state, there must be an outward energy flux of at least $9.5 \times 10^{-4} M_\odot\,{\rm km^2\, s^{-2} Myr^{-1}}$. The origin of this energy flow does not appear to be the hardening of binaries in the cluster core, at least as judged by the evidence from \gaia's renormalised unit weight error (RUWE). Rather the present-day mass loss of $0.26 M_\odot {\rm Myr}^{-1}$ due to tidal stripping by itself implies a substantial energy flow beyond the required magnitude. As the energy flow exceeds what is needed, the scale length of the low mass population will increase further as these stars are gradually subsumed into the Galaxy.

A consequence of mass segregation driven by two-body relaxation is the concentration of binaries in the high mass population in the centre. Of course, the hardening of binaries at the centres of globular clusters plays an important role in averting core collapse~\citep[e.g.,][]{He11}. In open clusters, simulations also suggest that hard binaries form~\cite[see Figure 11 of][]{Sp16}, though the importance of their role is less clear-cut. Radial velocity variability is an obvious probe of binarity, but such data are mainly available for the brighter, higher mass Hyades stars. {\gaia}'s renormalised unit weight error (RUWE) offers us a partial probe, as the astrometric trajectory of an unresolved binary has extra variance due to orbital motion. A high RUWE value can indicate that the source is an unresolved binary ~\citep[e.g.,][]{penoyre2020}, though it is still possible to hide hard binaries. If both the signal and measurement noise are large, then RUWE can still be small. Analysis of the RUWE distributions of the high and low mass populations suggests that their binary properties are rather similar, with if anything a longer tail to high RUWE values in the low mass population. Though tentative, this hints at a relative deficiency of binaries in the centre.

The Hyades (680 Myr) is an open cluster at the end of its life. It would be interesting to study mass segregation and binary populations in younger open clusters, such as the Pleiades \citep[115 Myr,][]{Ba96}. OE also indicated in their Figure 9 that the internal kinematics of Coma Berenices (560 Myr), IC 2602 (36 Myr) and Praesepe (625 Myr) are already resolvable with current \gaia\ astrometry~\citep[ages from][]{Si14}. Moving further afield, {\gaia}'s homogeneous photometry at the mmag level allows the open clusters in a radius of 4 kpc around our location to be studied systematically via colour-magnitude diagrams~\citep{cantat2018}. Variation of properties with Galactocentric radius is expected as the frailty of open clusters means they are susceptible to their habitat. Future \gaia\ data releases starting from DR3 will also provide a more direct characterization of binary populations in nearby open clusters, facilitating studies of the role of binaries in the dynamical evolution of clusters.

\section*{Acknowledgements}

SO is supported by the Science and Technology Facilities Council of the United Kingdom. This work has made use of data from the European Space Agency (ESA) mission Gaia (https://www.cosmos.esa. int/gaia), processed by the Gaia Data Processing and Analysis Consortium (DPAC, https://www.cosmos.esa. int/web/gaia/dpac/consortium). Funding for the DPAC has been provided by national institutions, in particular the institutions participating in the Gaia Multilateral Agreement. 
This paper made use of the Whole Sky Database (wsdb) created by Sergey
Koposov and maintained at the Institute of Astronomy, Cambridge by
Sergey Koposov, Vasily Belokurov and Wyn Evans with financial support
from the Science \& Technology Facilities Council (STFC) and the
European Research Council (ERC). We thank the referee for helpful comments on the manuscript.

%%%%%%%%%%%%%%%%%%%%%%%%%%%%%%%%%%%%%%%%%%%%%%%%%%

\section*{Data Availability}

The codes and data are available on request to the authors.



\bibliographystyle{mnras}
\bibliography{example} % if your bibtex file is called example.bib

%\begin{thebibliography}{99}
%\end{thebibliography}

%%%%%%%%%%%%%%%%%%%%%%%%%%%%%%%%%%%%%%%%%%%%%%%%%%

%%%%%%%%%%%%%%%%% APPENDICES %%%%%%%%%%%%%%%%%%%%%

%\appendix


%If you want to present additional material which would interrupt the flow of the main paper, it can be placed in an Appendix which appears after the list of references.

%%%%%%%%%%%%%%%%%%%%%%%%%%%%%%%%%%%%%%%%%%%%%%%%%%

% Don't change these lines
\bsp	% typesetting comment
\label{lastpage}
\end{document}